# Anomaly Detection and Inlet Pressure Prediction in Water Distribution Systems Using Machine Learning

*Tran Dang Khoa*

*Faculty of Computer Science and Engineering, Ho Chi Minh City University of Technology (HCMUT), 268 Ly Thuong Kiet St., Dist. 10, Ho Chi Minh City 740050, Vietnam*

ABTRACT

**Objective**: This study aims to develop two models to optimize pressure management in water distribution networks. The first model focuses on detecting anomalies in the water distribution system, such as leaks, blockages, or unusual pressure fluctuations. This model will predict future pressure at water distribution points and compare predicted values with actual values to assess anomalies. By doing so, the model will utilize the collected data to identify issues early, allowing managers to intervene promptly, thereby minimizing economic losses and safeguarding the sustainability of the water supply system. The second model is designed to predict the necessary inlet pressure based on the influence of distribution points in the system. Accurately predicting the inlet pressure ensures effective and continuous water supply for consumers, reduces waste, and improves water resource management. To achieve this, the model will apply modern machine learning algorithms to optimize the prediction process and analyze the factors affecting inlet pressure. This study not only enhances operational efficiency but also contributes to the development of sustainable solutions for the water distribution system in the context of climate change and increasing water demand.

**Methodology**: The study utilizes two main models to analyze and predict pressure in the water distribution network. The first model, CNN-EMD, is applied to analyze trends and predict future pressure values at water distribution points. Data on inlet pressure and pressure at distribution points are continuously collected over a two-month period with measurement frequency every 15 minutes. This model extracts important features from historical data by analyzing frequency components through the Empirical Mode Decomposition (EMD) method, helping to identify trends and pressure fluctuations, thus enabling early detection of anomalies in the system. The second model combines CNN, EMD, and LSTM to predict the necessary inlet pressure for the entire water network. This model focuses on analyzing the impact of distribution points on inlet pressure, thereby providing accurate predictions to ensure effective water supply.

**Results**: The study successfully applied two models, CNN-EMD and CNN-EMD-LSTM, to optimize pressure management in the water distribution network. The first model demonstrated high accuracy in predicting pressure at distribution points, identifying fluctuations and anomalies such as leaks and blockages with an accuracy detection rate ranging from 85% to 95%, depending on the distribution point. The use of the EMD method improved the ability to identify frequency components, thereby supporting more effective predictions of future pressure. The second model predicted inlet pressure with an average accuracy of 93%, allowing managers to adjust the system flexibly and effectively while identifying critical factors affecting inlet pressure.

**Conclusion**: The study demonstrates that applying advanced machine learning models like CNN-EMD and LSTM can significantly enhance pressure management capabilities in water distribution networks. The anomaly detection model helps identify issues early and provides information for timely intervention, thus minimizing economic damage and safeguarding the system's sustainability. The inlet pressure prediction model also contributes to ensuring effective water supply, reducing waste, and optimizing water resource management. The study's results have high practical applicability, contributing to improved water service quality and resource protection for future generations.

## 1 INTRODUCTION

**Inlet pressure management** is a crucial component of water distribution networks, directly impacting the stability and efficiency of distribution systems. Accurate prediction and control of inlet pressure are essential for maintaining optimal performance and ensuring reliable water delivery to consumers. However, traditional methods for predicting inlet pressure, which often rely on statistical techniques and manual adjustments, may struggle to fully capture the complexity and temporal dynamics of pressure fluctuations in modern water distribution systems.

Recent advancements in machine learning offer new opportunities to enhance the prediction and management of inlet pressure. **Convolutional Neural Networks (CNNs)** [2] and **Long Short-Term Memory (LSTM)** [3] networks have emerged as powerful tools for analyzing complex time series data and identifying intricate patterns that traditional methods might overlook. CNNs are particularly effective in extracting features from time series data by recognizing spatial patterns and trends [4], while LSTMs excel at capturing temporal dependencies and making predictions based on historical data.

*Author: Tel: +84 339 187 015*
*Email Address: khoa.tran107@hcmut.edu.vn*

This study investigates the integration of CNN-EMD and LSTM models to improve the prediction of inlet pressure in water distribution networks. We leverage CNNs to analyze the characteristics of inlet pressure data and combine them with LSTM to understand temporal dynamics, aiming for more accurate and reliable pressure predictions. The approach involves using CNN-EMD to process historical pressure measurements and data from various distribution points, followed by LSTM analysis to predict future pressure trends and make appropriate adjustments.

In addition to improving prediction accuracy, this integrated CNN-EMD model is also designed to detect anomalies within the water distribution network. Anomalies such as sudden drops or spikes in inlet pressure can indicate potential issues like leaks, blockages, or equipment failures. The model monitors the discrepancy between predicted and actual pressure values, allowing for the identification of significant deviations to detect these anomalies. This capability is crucial for proactive maintenance and timely intervention, helping to prevent minor issues from escalating into major problems.

The integration of CNN-EMD and LSTM models in this study represents a novel approach to addressing the challenges of inlet pressure management. The primary objective is to demonstrate the effectiveness of this combined model in enhancing prediction accuracy, anomaly detection, and operational efficiency. By comparing the performance of this model with traditional statistical methods and other machine learning techniques, we aim to highlight the potential benefits of advanced neural network models in improving water distribution network management.

## 2 RELATED WORKS

### 2.1 Traditional Methods in Inlet Pressure Management

**Previous studies** have primarily relied on traditional methods for managing and predicting inlet pressure, which often involve statistical techniques and physical models. These methods include **linear regression** [5]**, ARIMA models** [6]**, and feedback control techniques** [7] for regulating pressure in water distribution networks. However, these approaches often fall short in addressing the complexities of modern water distribution systems, especially when it comes to predicting intricate pressure fluctuations.

### 2.2 Machine Learning Applications in Water Distribution Systems

**Recently,** research has increasingly applied machine learning in managing water distribution networks. Several studies have employed machine learning algorithms such **as Support Vector Machines (SVM)** [8]**, Random Forests** [9]**, and Artificial Neural Networks (ANN)** [10] to predict pressure and detect anomalies. While these studies have demonstrated the effectiveness of machine learning in improving pressure prediction accuracy, there remain limitations in handling the complex spatial and temporal relationships in the data.

### 2.3 Convolutional Neural Networks (CNN) in Time Series Analysis

**CNNs** have been widely used across various fields, including image processing and pattern recognition. In time series analysis, CNNs have been employed to extract spatial features from time series data, particularly when data patterns exhibit spatial structure or trends. However, the application of CNNs in analyzing pressure data within water distribution networks is still limited and warrants further investigation.

### 2.4 Long Short-Term Memory (LSTM) Networks for Temporal Dynamics

**LSTM** is one of the most effective deep learning models for capturing long-term temporal dependencies. Research has applied LSTM to numerous time series prediction tasks, including weather forecasting, financial predictions, and energy management. LSTM is particularly effective in handling time series with continuous and unpredictable changes. In the context of water distribution networks, LSTM has been tested for predicting flow and pressure, but there is limited research on combining it with other methods to enhance accuracy.

### 2.5 Hybrid Models Combining CNN and LSTM

**Recent studies** have explored the combination of CNN and LSTM to leverage the strengths of both models. CNNs are used to extract features from the input data, while LSTMs capture the temporal relationships from these features. These hybrid models have shown superior performance in various applications, such as financial analysis, energy demand forecasting, and anomaly detection [11]. However, the application of such combined models in managing inlet pressure in water distribution networks has not been fully explored and requires further research.

### 2.6 Anomaly Detection in Water Distribution Networks

**Anomaly detection in water distribution networks** is a critical area of research. Traditional methods often rely on threshold-based approaches or simple statistical analysis to identify anomalies. However, with the advancement of machine learning, deep learning-based anomaly detection methods have been proposed, including using predictive models that monitor deviations between predicted and actual values. Some studies suggest that combining CNN and LSTM for anomaly detection can achieve higher accuracy, especially in systems with complex temporal variations.

## 3 IMPLEMENTATION

### 3.1 Data Collection and Handling Missing Values:

**3.1.1 Data Collection:**

For this study, we collected data on pressure and flow rates from the water distribution system, including measurements taken from all relevant points within the system. The dataset consists of inlet pressure and flow rate readings, recorded at 15-minute intervals over a period of 2 months. This data was sourced from the SCADA (Supervisory Control and Data Acquisition) system, which provides real-time monitoring and data acquisition, ensuring comprehensive and accurate measurements of pressure and flow throughout the water distribution system.

*Author: Tel: +84 339 187 015*
*Email Address: khoa.tran107@hcmut.edu.vn*

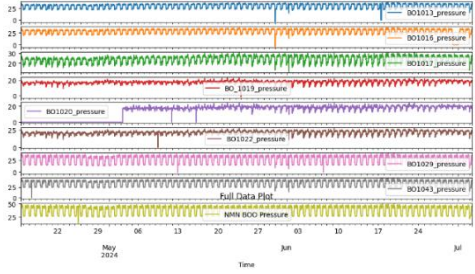

**Fig. 1:** Pressure measurements collected every 15 minutes over 2 months from the SCADA system.

### 3.1.2 Handling Missing Values:

In this phase, we address missing and invalid values using random forest regression:

- **Data Conversion:** The Date Time column is converted to a datetime format, enabling the extraction of time-based features such as hour and minute, which are essential for the model.

- **Model Training:** We filter out valid pressure values to train the random forest regression model. This model learns the relationship between time features and pressure readings.

- **Prediction and Imputation:** For entries with invalid or missing pressure values, the trained model predicts pressure values based on the time features. These predicted values are then used to replace the invalid entries, ensuring a more accurate and complete dataset.

### 3.1.3 Advantages:

- **Improved Data Quality:** This method enhances data quality by replacing missing or invalid values with reasonable predictions, resulting in a more complete and accurate dataset.

- **Increased Analysis Reliability:** Clean and accurate data improves the reliability of subsequent analyses and predictive models.

- **Ease of Application:** Random forest is a powerful yet easy-to-apply machine learning method for prediction and data handling. It provides robust results that are both reliable and interpretable, making it suitable for complex datasets.

- **Utilization of Time Features:** By incorporating time-based features, the model can capture temporal trends and patterns in the data, leading to more accurate predictions.

This approach not only improves data quality but also supports more precise analysis and predictions in later stages.

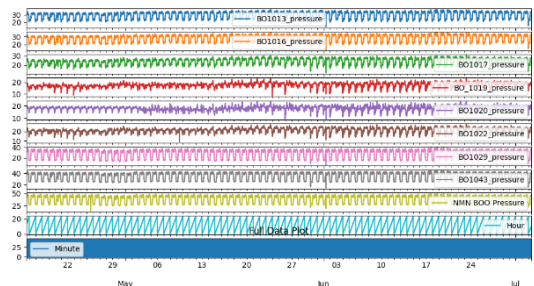

**Fig. 2:** Data after handling missing values with random forest regression predictions.

*Author: Tel: +84 339 187 015*
*Email Address: khoa.tran107@hcmut.edu.vn*

### 3.2 Preprocessing

In this study, the data preprocessing method includes the main steps of **data extraction, data normalization, One-Hot Encoding of temporal features and storage of normalization parameters.** The aim of the preprocessing process is to prepare the input and output data for analytical models, ensuring consistency and improving model performance.

#### 3.2.1 Data Extraction

- **Target Variable Extraction:** The primary focus of this study was to predict the inlet pressure. Therefore, the target variable, which is the specific inlet pressure values, was extracted from the DataFrame. This variable is crucial for the analysis as it represents the key outcome we aim to predict.

- **Feature Variables Extraction:** The feature variables are those that are used to predict the target variable. In this context, these include the pressure measurements and additional characteristics collected from various points in the network, excluding the target variable and time-related information. The feature variables were isolated from the DataFrame to prepare them for further processing.

- **Time Handling:** Time-related information, essential for understanding temporal patterns in the data, was extracted and converted into a suitable format. This included parsing the date and time fields to facilitate time-based analysis and preprocessing.

Overall, the extraction process ensured that the dataset was prepared with a clear separation between target and feature variables, and time information was effectively handled to support subsequent analysis and modeling.

#### 3.2.2 Data Normalization

In this study, normalization is applied to both input features and the target variable to ensure that all values are scaled to a consistent range. The normalization process is carried out using the Min-Max normalization method, which rescales the data to a fixed range, typically [0, 1].

- **Min-Max Normalization Method**: The Min-Max normalization technique is used to transform data values to fit within a specified range, usually between 0 and 1. This method is defined by the formula:

$$X_{norm} = \frac{X - X_{min}}{X_{max} - X_{min}}$$

  **X** is the original value,

  $X_{min}$ the minimum value in the dataset,

  $X_{max}$ is the maximum value in the dataset,

  $X_{norm}$ is the normalized value.

- **Normalization of Input Features**: The input features (X) are normalized using the Min-Max normalization method. This technique transforms the values of the features to fall within a specified range, usually [0, 1]. This ensures that each feature contributes equally to the model's learning process, preventing features with larger value ranges from dominating the learning process.

- **Normalization of Target Variable:** Similarly, the target variable (Y) is also normalized using the Min-Max method. The target values are scaled to the same range as the input features, ensuring consistency in the scale of variables. This consistency is crucial for effective model training and prediction.

### 3.2.3 One-Hot Encoding of Temporal Features

In this study, one-hot encoding is applied to temporal features to better capture the cyclical nature of time-based data. Features such as day, hour, and minute are converted into binary vectors through one-hot encoding. This technique transforms categorical variables—in this case, discrete time components—into a format that machine learning models can readily utilize. One-hot encoding ensures that these temporal variables are treated as independent categories, allowing the model to recognize and learn complex patterns related to specific times of day without implying any ordinal relationship between different time values.

**One-Hot Encoding**: One-hot encoding converts a categorical variable with nnn possible values into nnn binary features, each representing the presence or absence of a particular category. For example, if the "hour" feature has values ranging from 0 to 23, one-hot encoding will create 24 binary features, with one feature being "on" (1) and the rest "off" (0) for any given hour.

**Advantages**:

- **Non-ordinal Nature**: One-hot encoding helps in representing the temporal features as non-ordinal categories, ensuring that the model does not interpret them as having an inherent order or progression, which is crucial for features like "day" or "hour."
- **Cyclical Time Representation**: By encoding time components as separate binary features, the model can better capture patterns that are cyclical, such as hourly or daily trends, which are common in time series data.

One-hot encoding of temporal features allows the model to effectively learn patterns associated with different times of the day or month without introducing unintended ordinal relationships.

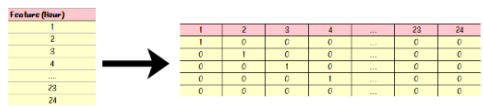

**Fig. 3:** Illustration of the One-Hot Encoding Process for the Hour Feature

### 3.2.4 Storage of Normalization Parameters

To ensure that the model can be applied consistently to new data and maintain the integrity of its predictions, the normalization parameters used during the preprocessing stage are saved. These parameters include the minimum and maximum values of each feature (or target variable) that were calculated during the Min-Max normalization process.

By storing these parameters, the same scaling transformation can be applied to any future data, ensuring consistency between the training and inference phases. This step is crucial for models that require input data to be normalized in the same way during both training and testing.

In practice, the normalization parameters are saved to a file (e.g., in JSON format), allowing them to be easily reloaded and reused. This approach avoids recalculating the parameters from the original dataset, which might not always be available when deploying the model for prediction.

## 3.3 Predict Future Pressure Data Using CNN

### 3.3.1 Autocorrelation and Partial Autocorrelation Analysis

The Autocorrelation Function (ACF) and Partial Autocorrelation Function (PACF) [12] are essential tools in time series analysis, both of which help identify underlying patterns in the data and aid in building predictive models.

- **ACF (Autocorrelation Function):** ACF measures the correlation between the values of a time series with its own past values at different lags. This helps determine whether previous values influence the current value and how strong that influence is. ACF analysis is commonly used to check for stationarity in time series data and to identify the structure of the autoregressive (AR) component in a model.
- **PACF (Partial Autocorrelation Function):** PACF measures the correlation at a specific lag after removing the effects of all previous lags. PACF is useful in determining the number of lags that directly relate to the current value of the time series, which helps in building more accurate forecasting models.

By combining ACF and PACF analysis, one can detect autoregressive patterns and dependencies in the time series, leading to the selection of the most appropriate model for analyzing and forecasting the data.Bottom of Form

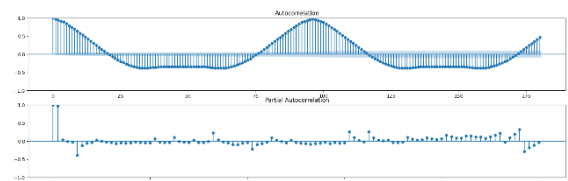

**Fig. 4:** ACF and PACF Analysis of Inlet Pressure Data

### 3.3.2 Empirical Mode Decomposition (EMD)

Empirical Mode Decomposition (EMD) [13] is a powerful nonlinear analysis technique used to transform nonstationary and nonlinear time series data into stationary and linear components. Unlike traditional decomposition methods such as Fourier [14] or wavelet decomposition [15], EMD offers an adaptive approach that aligns with the intrinsic time scales of the data. This adaptability allows EMD to decompose a time series into intrinsic mode functions (IMFs) and a residual, capturing the data's local features across different time scales.

*Author: Tel: +84 339 187 015*
*Email Address: khoa.tran107@hcmut.edu.vn*

**Key Characteristics of EMD**

EMD is characterized by the following properties for each intrinsic mode function:

- **Extreme Points and Zero-Crossings:** The number of extreme points (local maxima and minima) in each IMF differs from the number of zero-crossings by no more than one.

- **Envelope Mean:** The average of the upper and lower envelopes of each IMF must be zero.

**Execution Process of EMD**

The EMD algorithm follows a systematic process to decompose a time series:

1. **Identify Local Extrema:** Determine all local maxima and minima in the time series $y(t)$.

2. **Generate Envelopes:** Construct the upper envelope $y_u(t)$ and lower envelope $y_l(t)$ using cubic spline interpolation based on the identified extrema.

3. **Compute Mean Envelope:** Calculate the mean of the upper and lower envelopes:

$$m(t) = \frac{y_u(t) + y_l(t)}{2}$$

4. **Extract Intrinsic Mode Function (IMF):** Determine the difference between the original time series and the mean envelope:

$$d(t) = y(t) - m(t)$$

5. **Verify IMF Properties:** Check if the candidate IMF $d(t)$ meets the following criteria:

   • The number of zero-crossings and extrema should not differ by more than one.

   • The average of the upper and lower envelopes should be zero.

   If $d(t)$ meets these criteria, it is considered a valid IMF. Otherwise, replace $y(t)$ with $d(t)$ and repeat steps 1 to 4.

6. **Extract Residual:** Once a valid IMF is obtained, calculate the residual $r(t)$ as:

$$r(t) = y(t) - d(t)$$

Use this residual as the new time series and return to step 1. Continue the process until the residual becomes a monotonic trend or is sufficiently smooth.

7. **Reconstruct the Original Series:** After obtaining all IMFs and the residual, reconstruct the original time series y(t) by summing all the IMFs and the residual:

$$y(t) = r(t) + \sum_{i=1}^{n} d_i(t)$$

where $d_i(t)$ represents the i-th IMF and n is the number of IMFs extracted.

EMD effectively decomposes a time series into intrinsic mode functions that capture different time scales and trends, providing valuable insights for data analysis and forecasting.

In this study, the Empirical Mode Decomposition (EMD) method is applied to analyze inlet pressure data. By utilizing EMD, the inlet pressure data is decomposed into Intrinsic Mode Functions (IMFs) and a residual component. Each IMF captures the characteristics of the data at different timescales, enabling the machine learning model to better grasp important features in the inlet pressure data.

Once the inlet pressure data is decomposed by EMD, the IMF components are normalized and formatted into a matrix suitable for input into a Convolutional Neural Network (CNN) model. The matrix is structured as follows:

$$X_M(t) = \begin{pmatrix} imf_1(t-1) & imf_1(t-2) & \cdots & imf_1(t-29228) \\ imf_2(t-1) & imf_2(t-2) & \cdots & imf_2(t-29228) \\ \vdots & \vdots & \ddots & \vdots \\ imf_C(t-1) & imf_C(t-2) & \cdots & imf_C(t-29228) \end{pmatrix}$$

Where t represents the current time, and $imf_C(t-i)$ denotes the value of the C-th IMF component after normalization, recorded i time intervals before the current time t. The matrix contains data spanning a 10-month period, with values recorded every 15 minutes. The dimensions of the matrix are (8, 29228), representing 8 IMF components and 29228, 15-minute intervals. This matrix is used to train the CNN model, improving its ability to analyze and predict pressure fluctuations in the water supply network with greater accuracy.

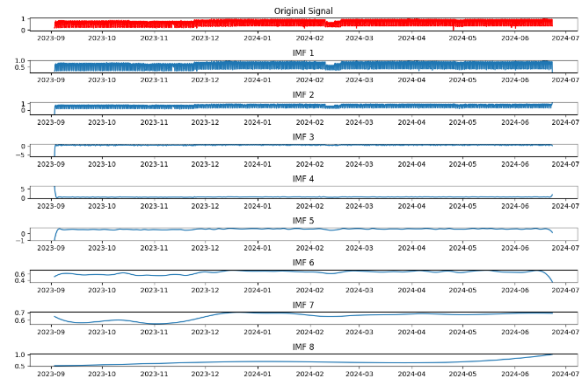

**Fig. 5:** Illustration of IMF Components Analyzed After the EMD Process

### 3.3.2 Hilbert-Huang Transform

In this study, the Hilbert-Huang Transform (HHT) [16] is applied to analyze the instantaneous frequency characteristics of inlet pressure data, providing deeper insights into the complex fluctuations within the water distribution network. HHT is a powerful method for separating various frequency components in data, especially effective in handling non-linear and non-stationary signals.

*Author: Tel: +84 339 187 015*
*Email Address: khoa.tran107@hcmut.edu.vn*

1. **Application of Hilbert Transform**

After extracting the Intrinsic Mode Functions (IMFs) from the inlet pressure data, the Hilbert Transform is employed to analyze the instantaneous properties of each IMF. This process includes calculating the instantaneous phase of each IMF, allowing us to detect subtle shifts in the frequency of the pressure signal over time. This capability of HHT distinguishes it from traditional methods like Fourier transforms, which cannot capture these instantaneous variations.

**2 Instantaneous Frequency Analysis**

The instantaneous frequency analysis provides detailed insights into how the frequency components within the data change over time. This can reveal minor anomalies in the inlet pressure as well as long-term trends associated with varying operational conditions within the water distribution network. By capturing these subtle and unusual changes, the system can respond quickly to maintain stability in network operations.

**3 Contribution to Forecasting and Control Models**

The application of HHT significantly enhances the model's ability to forecast pressure changes and detect potential anomalies. By integrating this analysis into the control system, the management of the water distribution network is optimized, improving responsiveness and efficiency in real-world situations. These improvements not only ensure the stability of the network but also help to optimize water distribution, thereby enhancing service quality and reducing the risks of unexpected incidents.

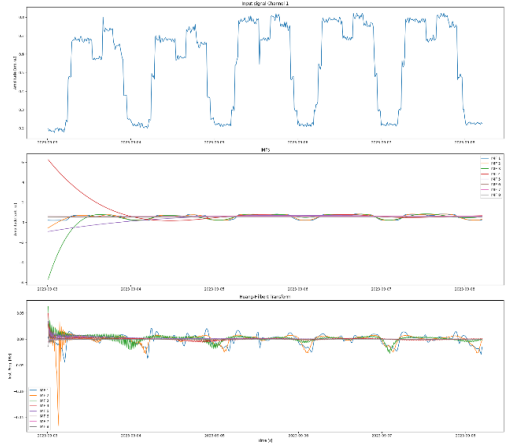

**Fig. 6:** Comprehensive Hilbert-Huang Transform Analysis of the Full Time Series

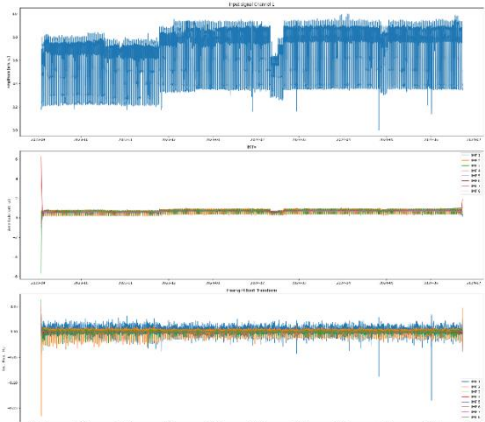

**Fig. 7**: Comprehensive Hilbert-Huang Transform Analysis of 500 Data Points

*Author: Tel: +84 339 187 015*
*Email Address: khoa.tran107@hcmut.edu.vn*

### 3.3.2 Series to Supervised Learning

In time series forecasting, converting time series data into a supervised learning [17] format is crucial to leveraging machine learning models. This technique, known as Series to Supervised Learning, transforms sequential time-based data into input-output pairs, making it suitable for prediction tasks.

The primary approach involves shifting time-series data to create lagged features, where past observations are used as predictors for future values. For instance, given a time series with observations at each time step t, the values at time t-1, t-2, ..., are used as input features to predict the target value at time t. This transformation results in a structured dataset, where each row contains the past observations and the corresponding target value.

By creating new columns for each lag, the continuous time series data is converted into a supervised learning format. This allows traditional machine learning models, such as linear regression, decision trees, or neural networks, to be applied effectively. Specifically, the lagged features capture the temporal dependencies, which are essential for accurate forecasting.

The Series to Supervised Learning approach is particularly beneficial for time series prediction tasks due to the following advantages:

- Capturing Temporal Relationships: By leveraging past values, models can learn and exploit temporal patterns in the data, improving predictive accuracy.

- Versatility: The transformed dataset can be utilized by a wide range of machine learning models, from simple linear models to advanced deep learning architectures like LSTM and CNN.

This method is commonly employed in various forecasting applications, including energy demand prediction, stock price forecasting, and other temporal data analysis tasks. Converting the time series data into a supervised format enables models to effectively learn from past trends and make more informed predictions about future events.

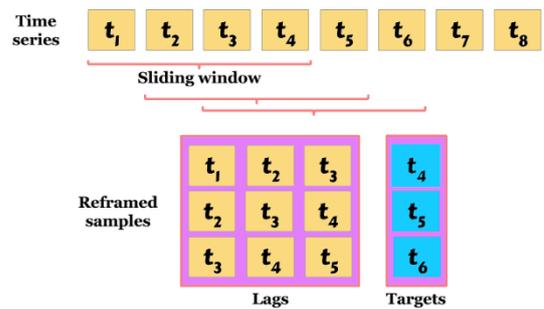

**Fig. 8:** Illustration of the Series to Supervised Learning Process

### 3.3.3 Training the CNN Model with IMF Data.

In this study, we employ a 1D Convolutional Neural Network (CNN) to train a predictive model based on Intrinsic Mode Functions (IMFs) derived from Empirical Mode Decomposition (EMD). The following are the key steps involved in the CNN training process, along with techniques and methods to optimize model performance:

**1. Building the CNN Model**

The CNN model is designed with multiple 1D convolutional layers to extract complex features from the IMF data. The architecture includes several convolutional layers with varying configurations, such as filter size, kernel size, dilation rate, and activation functions. The model also incorporates optional BatchNormalization layers to improve training stability and reduce overfitting.

- **Layer Structure:** The model consists of several Conv1D layers with causal padding, enabling it to capture temporal patterns in the time-series data. Different dilation rates are used in successive layers to capture both short-term and long-term dependencies.

- **Feature Extraction:** The convolutional layers operate in parallel to extract diverse features from the input IMFs, which are then combined to form a comprehensive feature set for prediction.

- **Final Prediction:** The extracted features are passed through additional convolutional layers to generate the final prediction.

**2. Compiling and Training the Model**

After building the CNN model, it is compiled using the Adam optimizer and the Mean Squared Error (MSE) loss function. Key aspects of the training process include:

- **Optimization:** The learning rate is dynamically adjusted to optimize the loss function, ensuring efficient model convergence.

- **Training Process:** The model is trained on the IMF data for a predefined number of epochs, with early stopping and learning rate reduction callbacks enhancing performance. The best model weights are saved during training.

**3. Evaluation and Prediction**

Once trained, the model is evaluated on validation data to ensure generalization. The trained model is then used to make predictions on new data, leveraging the features learned from IMF analysis.

This approach effectively harnesses the feature extraction capabilities of CNNs to improve predictive accuracy, particularly for complex time-series data such as water distribution network pressure data. By incorporating IMFs, the model captures intricate patterns, leading to more accurate forecasts and more efficient network management.

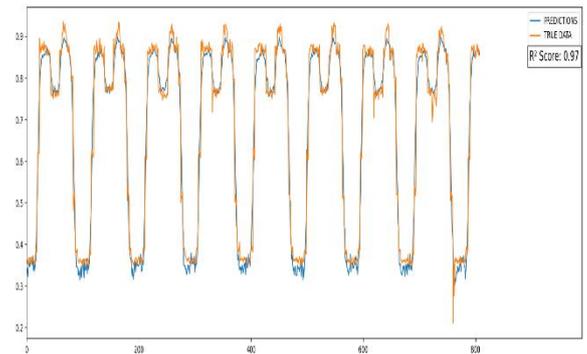

**Fig. 10:** Prediction Results with a Simple CNN Model

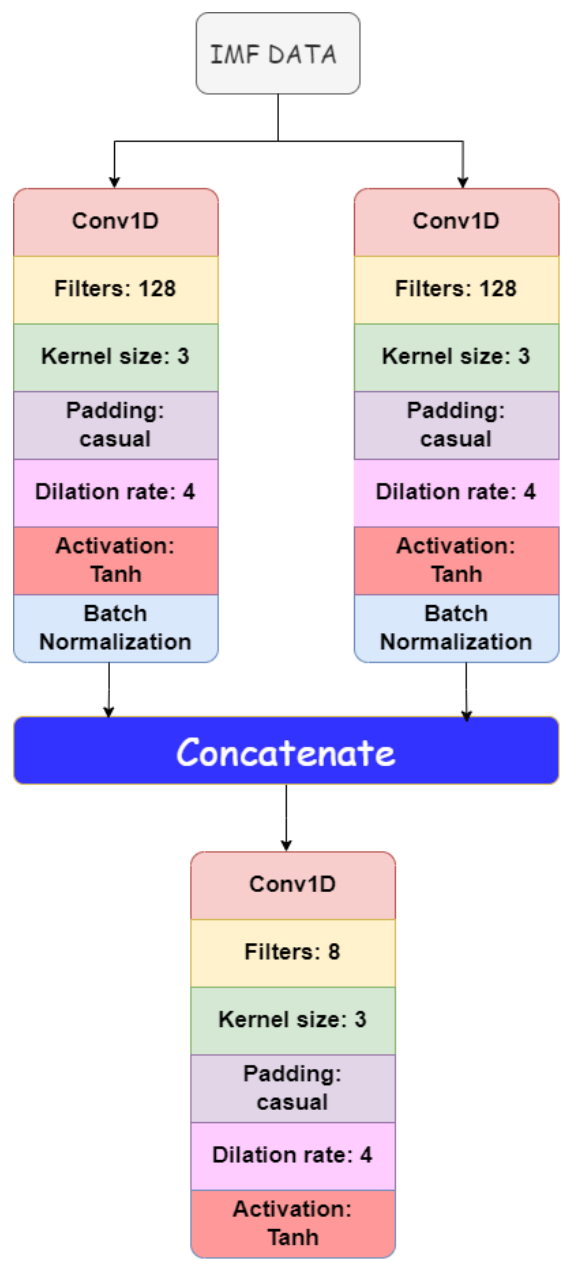

**Fig. 9:** Training IMF Data with a Simple CNN Model

*Author: Tel: +84 339 187 015*
*Email Address: khoa.tran107@hcmut.edu.vn*

*3.4 Training the Model with an LSTM Network*

In this study, we apply an integrated approach combining Convolutional Neural Networks (CNN) and Long Short-Term Memory (LSTM) networks to predict inlet pressure in water distribution networks. Below is a detailed description of how we use the input datasets and the training process of the model.

**3.4.1 Input Data and Processing**

**1. Pressure Data at Distribution Points:** This dataset provides information on water pressure at various distribution points in the system. To predict future values, the data is processed in a "series to supervised" format with a lookback window of 96, corresponding to 24 hours (15-minute intervals x 96). This setup allows the LSTM model to learn and predict trends in pressure fluctuations within the system.

**2. Flow Data at Distribution Points:** Similarly, flow data is processed with a lookback window of 96. This enables the LSTM model to learn and analyze changes in water flow over time to predict future flow trends. Flow data provides a comprehensive view of fluctuations in the water distribution system.

**3. Inlet Pressure Data:** This third input is processed through a CNN to predict pressure values at the next time step. The CNN is designed to capture spatial and temporal features from the inlet pressure data, providing useful information for subsequent predictions. The predictions from the CNN are then fed into an LSTM model for further analysis.

**3.4.2. Data Integration and Synthesis Process**

**1. Processing with CNN and LSTM:**

- **Pressure and Flow Data:** Processed using LSTM models. The LSTM model learns temporal relationships from the pressure and flow data, analyzing features and fluctuations within the system.
- **Inlet Pressure Data:** Processed through CNN to predict future pressure values. The CNN-generated predictions are then input into another LSTM model for deeper analysis.

**2. Combining Outputs**

After processing the data with the models, three outputs (pressure data from LSTM, flow data from LSTM, and inlet pressure predictions from CNN) are concatenated. This combination creates a synthesized representation from the features learned by each model, allowing the final model to learn from different information sources and improve prediction accuracy.

**3. Combined LSTM Model**

The concatenated outputs are fed into a final LSTM model. This LSTM model learns relationships between the processed inputs and synthesizes information to provide the most accurate predictions. The model continues to learn complex patterns and relationships in the data, enhancing prediction accuracy for the water distribution network.

**3.4.3. Impact on Network Control**

**1. Understanding Inlet Pressure Fluctuations**

Using CNN to predict future inlet pressure adds an additional layer of understanding to the model. By predicting inlet pressure and feeding these predictions into the LSTM model, the system can forecast pressure changes, allowing for more proactive adjustments to the water distribution network, rather than just reacting to current conditions.

**2. Optimizing Network Control**

This approach helps the model understand causal relationships between distribution point behavior and changes in inlet pressure. With this understanding, the model can optimize network control, improving water distribution efficiency and reducing operational costs. This methodology significantly enhances the management and control of the entire water distribution system.

This approach not only improves the accuracy of predicting key factors but also provides deeper insights into the dynamics and interactions within the water distribution network, thereby enhancing system effectiveness and reliability.

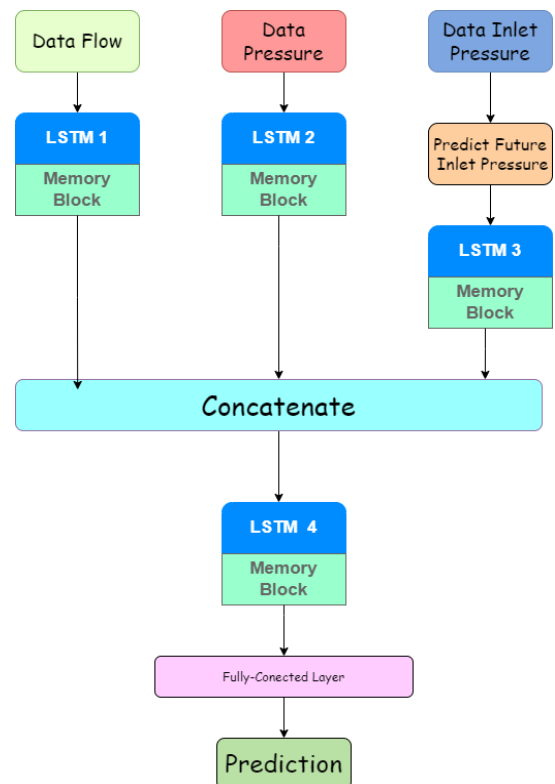

**Fig. 11:** Training a Simple CNN-LSTM Model

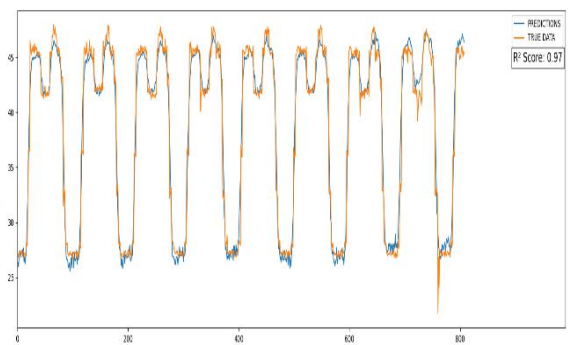

**Fig. 11:** Prediction Results with a Simple CNN-LSTM Mod

*Author: Tel: +84 339 187 015*
*Email Address: khoa.tran107@hcmut.edu.vn*

## 4 OVERVIEW PROJECT

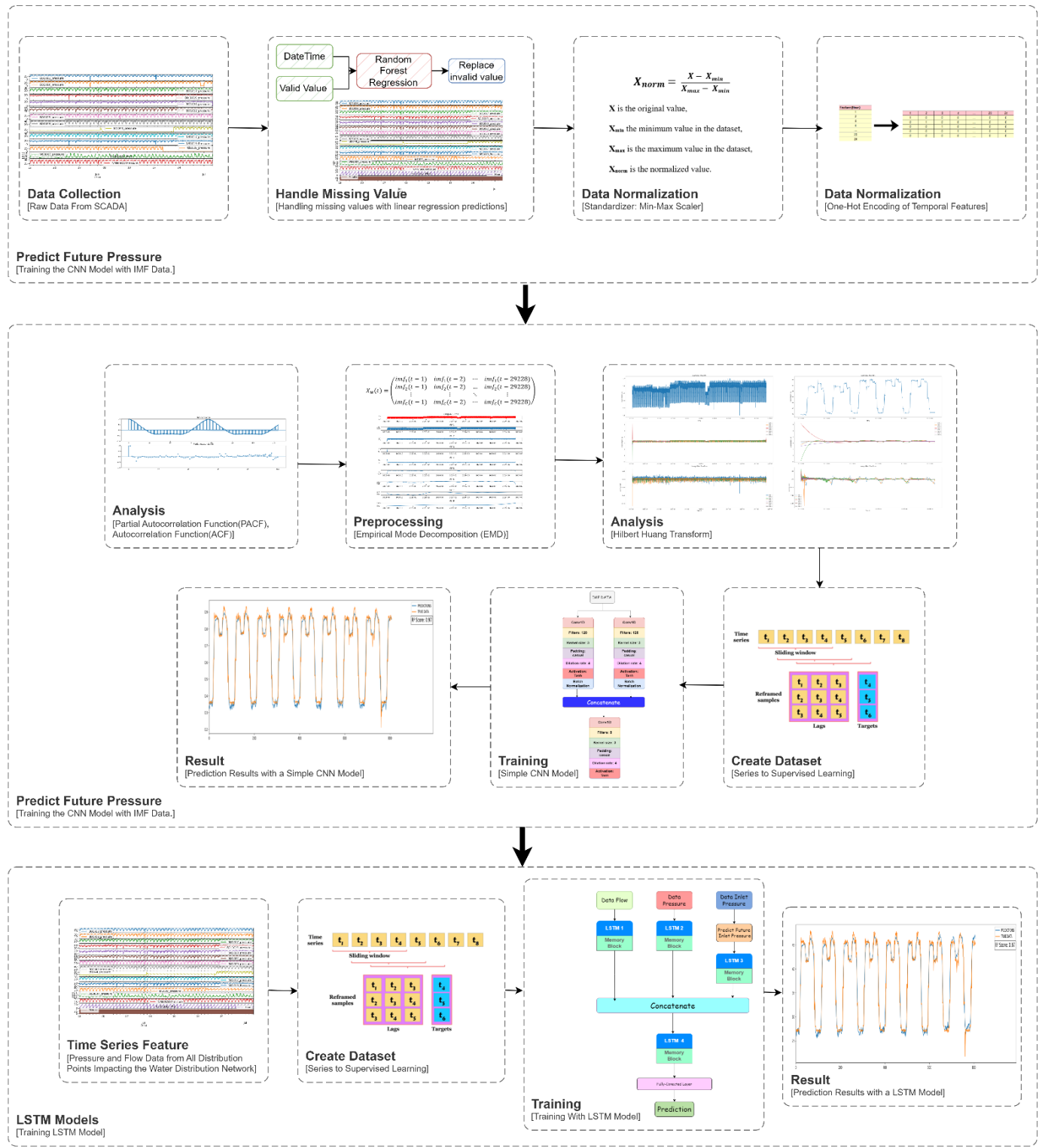

Fig. 12: Overview Project

*Author: Tel: +84 339 187 015*
*Email Address: khoa.tran107@hcmut.edu.vn*

# 5 RESULT PREDICT FUTURE.

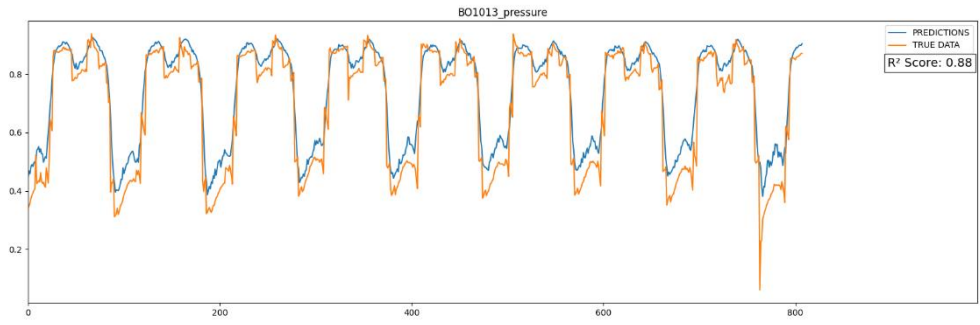

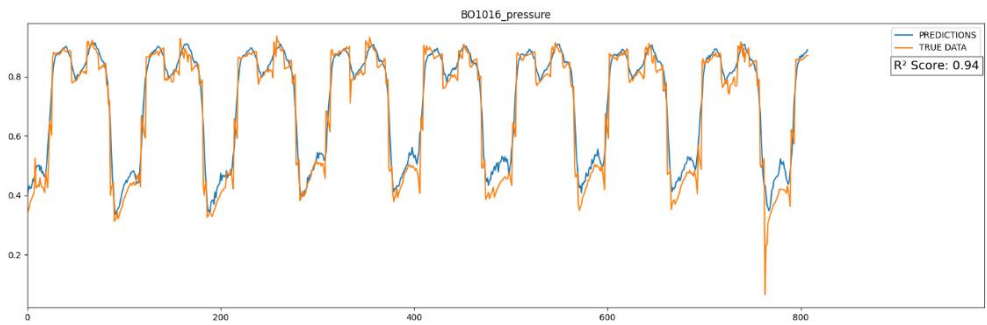

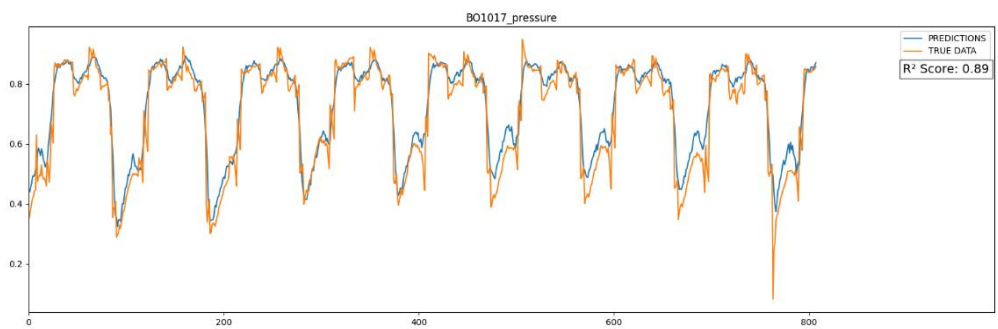

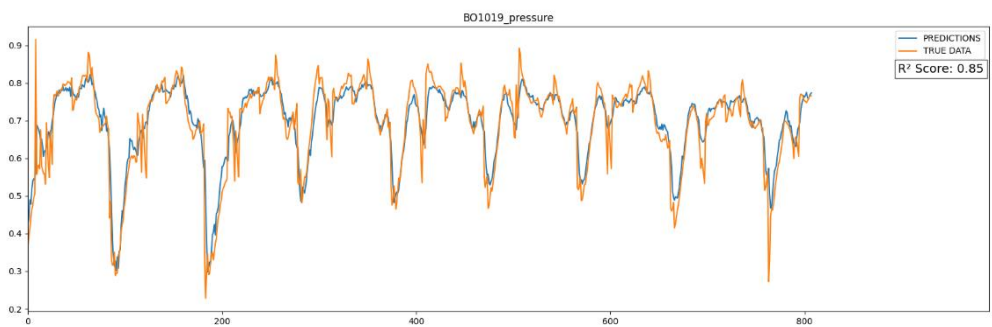

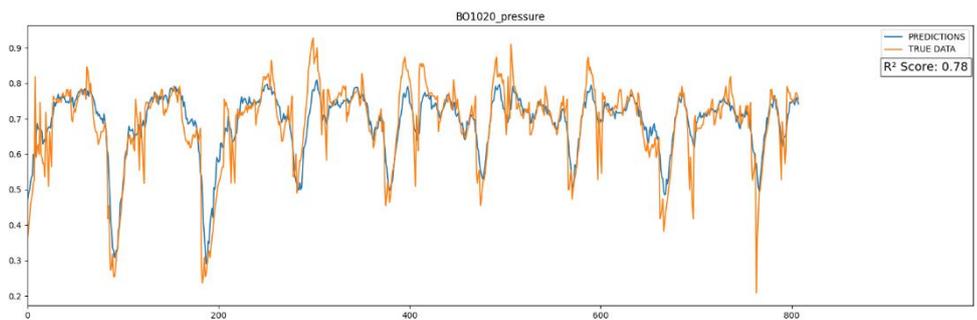

*Author: Tel: +84 339 187 015*
*Email Address: khoa.tran107@hcmut.edu.vn*

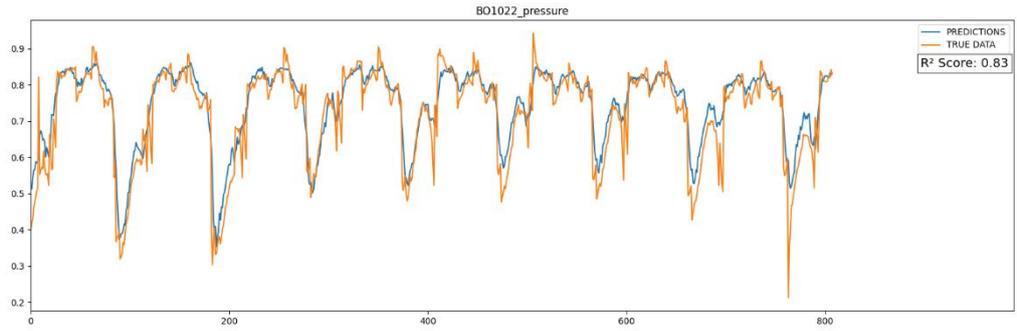

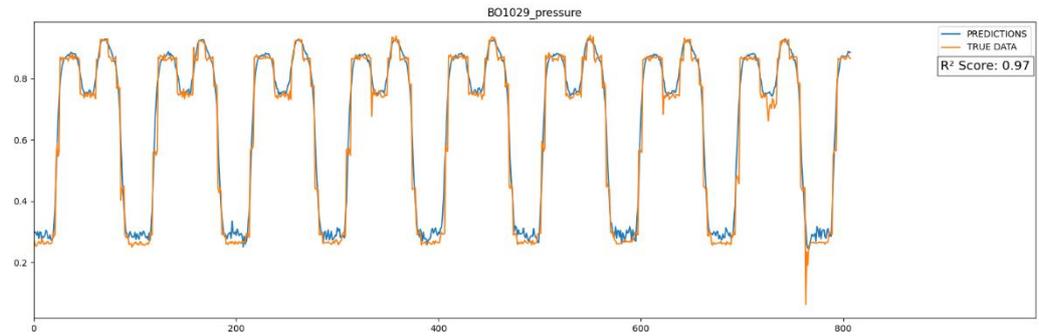

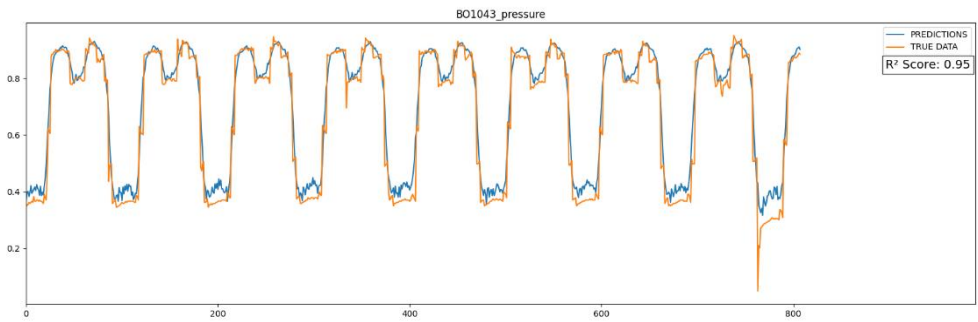

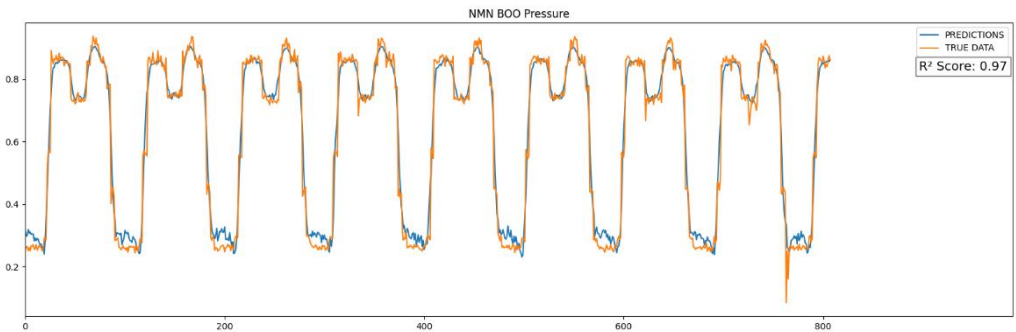

*Author: Tel: +84 339 187 015*
*Email Address: khoa.tran107@hcmut.edu.vn*

# 6  EXPERIENCE

*5.1 Handling Missing Data with New Data Points Lacking Historical Information*

**In dynamic systems** like water distribution networks, new data points often emerge without sufficient historical data for training predictive models. In such cases, special methods must be applied to integrate these new points into the predictive model. The approach involves the following steps:

1. **Creating a Small Model**: When a new data point appears without historical data, a small model is created to analyze the interaction between the new point and the existing data points in the system. This small model is used to assess how changes in the new point's value affect the existing points. This allows for a better understanding of the dynamic relationships between the new and old data points, providing a more comprehensive view of the system.

2. **Calculating the Impact**: The small model focuses on calculating the impact of the new data point on the system. Specifically, it observes how changes in the new point influence the behavior of the old data points that have already been incorporated into the trained model. From this, the model can assess the new point's influence on the overall system behavior, offering valuable insights into its role within the network.

3. **Incorporating Interpolation and Imputation Methods**: Based on the assumption that the new and existing data points follow the same underlying pattern, interpolation methods can be used to estimate values for the missing data points. Additionally, imputation techniques can be applied, using probabilistic distributions to fill in the missing values with the most likely estimates. This method not only provides predicted values but also includes associated probabilities, offering a level of confidence for each new data point's prediction.

4. **Integrating New Data into the Main Model**: Once the impact of the new data point has been understood through the small model, this information can be incorporated into the main predictive model. By integrating the influence of the new data point, the predictive model can make more accurate forecasts, even when lacking initial historical data. This approach allows the model to continuously improve and adapt as new data becomes available, enhancing the overall predictive performance of the system.

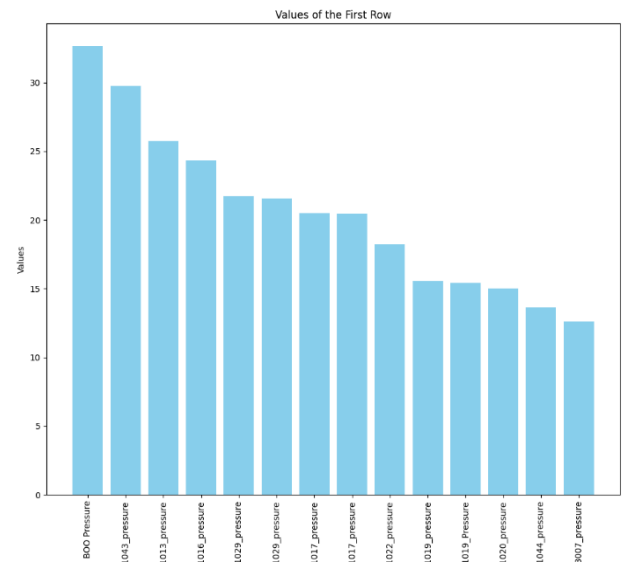

**Fig. 13**: Analyzing Pressure Trends in the Water Distribution Network

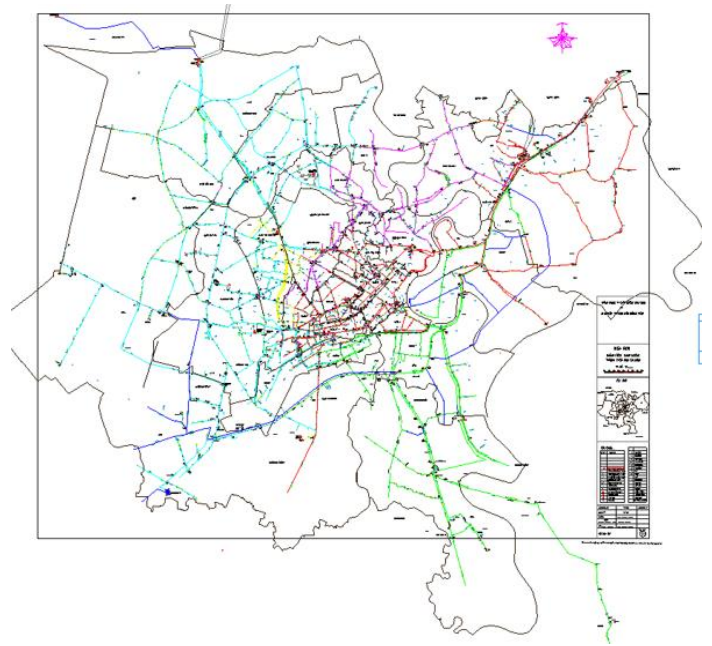

**Fig. 14**: Water Distribution Network in Ho Chi Minh City, Vietnam

*Author: Tel: +84 339 187 015*
*Email Address: khoa.tran107@hcmut.edu.vn*

*Author: Tel: +84 339 187 015*
*Email Address: khoa.tran107@hcmut.edu.vn*